\documentclass[letterpaper,leqno,11pt,oneside]{amsart}

\usepackage{graphicx}
\usepackage{amssymb}
\usepackage{amsthm}
\usepackage{bm}
\usepackage{amsmath}
\usepackage{mathrsfs}
\usepackage{yfonts}
\usepackage{enumerate}
\usepackage{lscape}
\usepackage{psfrag}
\usepackage{setspace}
\usepackage{hyperref}

 \usepackage[usenames]{color}
 \usepackage{comment}

\theoremstyle{plain}

\theoremstyle{definition} 

\def\beqs{\begin{eqnarray*}}
\def\eeqs{\end{eqnarray*}}

\def \Z {\mathbb Z}

\def \R {\mathbb R}

\def\be{\begin{equation}}
\def\ee{\end{equation}}

\newcommand{\EE}{\ensuremath{\mathbb{E}}}

\newcommand{\e}{\epsilon}
\newcommand{\p}{\partial}

\newcommand{\ep}{\epsilon}

\newcommand{\pp}{\mathbb{P}}

\newcommand{\rr}{\mathbb{R}}

\newcommand{\aipo}{\mathcal{A}_1}

\newcommand{\uno}[1]{\mathbf{1}_{#1}}

\newcommand{\wt}{\widetilde}

\newcommand{\K}{K_{\Ai}}

\DeclareMathOperator{\Ai}{Ai}

\renewcommand{\mu}{K}
\renewcommand{\nu}{M}

\begin{document}
\title[Renormalization fixed point of the KPZ universality class]{Renormalization fixed point of the KPZ universality class}

\author[I. Corwin]{Ivan Corwin}
\address{I. Corwin, Columbia University,
Department of Mathematics,
2990 Broadway,
New York, NY 10027, USA,
and Clay Mathematics Institute, 10 Memorial Blvd. Suite 902, Providence, RI 02903, USA,
and Massachusetts Institute of Technology,
Department of Mathematics,
77 Massachusetts Avenue, Cambridge, MA 02139-4307, USA
and Institut Henri Poincar\'{e},
11 Rue Pierre et Marie Curie, 75005 Paris, France}
\email{ivan.corwin@gmail.com}

\author[J. Quastel]{Jeremy Quastel}
\address{J. Quastel\\
  Department of Mathematics\\
  University of Toronto\\
  40 St. George street\\
  Toronto Ontario, Canada M5S 2E4.}
\email{quastel@math.utoronto.ca}

\author[D.~Remenik]{Daniel Remenik}
\address{D.~Remenik\\
  Departamento de Ingenier\'ia Matem\'atica and Centro de Modelamiento Matem\'atico\\
  Universidad de Chile\\
  Av. Beauchef 851, Torre Norte\\
  Santiago\\
  Chile} \email{dremenik@dim.uchile.cl}

\date{\today}

\begin{abstract}
The one dimensional Kardar-Parisi-Zhang universality class is believed to describe many types of evolving interfaces which have the same characteristic scaling exponents. These exponents lead to a natural renormalization/rescaling  on the space of such evolving interfaces. We introduce and describe the renormalization  fixed point of the Kardar-Parisi-Zhang universality class in terms of a random nonlinear semigroup with stationary independent increments, and via a variational formula. Furthermore, we compute a plausible formula the exact transition probabilities using replica Bethe ansatz. The semigroup is constructed from the {\it Airy sheet}, a four parameter space-time field which is the Airy$_2$ process in each of its two spatial coordinates. Minimizing paths through this field describe the renormalization group fixed point of directed polymers in a random potential. At present, the results we provide do not have mathematically rigorous proofs, and they should at most be considered proposals.
\end{abstract}

\maketitle

\section{Introduction}

Interfaces evolving according to local stochastic growth rules have been extensively studied in physics, biology, material science, and engineering. There has been significant theoretical success over the last twenty-five years in describing such rough self-affine interfaces which evolve due to local processes. In short times such interfaces display properties particular to their local processes. However, in the long time limit, it is predicted that certain universal scaling exponents and exact statistical distributions should accurately describe the {\it fixed long time} properties of a wide variety of rough interfaces. These predictions have been repeatedly confirmed through Monte-Carlo simulation as well as experiments. What is lacking, however, is a theoretical understanding and prediction of the temporal evolution of these interfaces, in the large-time scaling limit. In this article we provide two complementary descriptions of this universal temporal evolution of interfaces -- one from a PDE perspective and the other from an exact solvability perspective. As of yet, neither of these descriptions are justified with mathematical proof. However, a variety of predictions arising from these perspectives can be rigorously confirmed (e.g. \cite{CFP1,cqr,QuastelRemenikReview,CH2,CorwinLiuWong}).

The $1+1$ dimensional KPZ universality class includes a wide variety of forms of
stochastic interface growth \cite{FS2, KS, BSt} on a one dimensional substrate, randomly
stirred one dimensional fluids (the stochastic Burgers equation) \cite{FNS}, polymer
chains directed in one dimension and fluctuating transversally in the other due to a
random potential \cite{HH} and various lattice models such as the driven lattice gas model
of ASEP and ground state polymer model of last passage percolation \cite{J}. All models
can be transformed to a kinetically roughening, growing interface reflecting the
competition between growth in a direction normal to the surface, a surface tension
smoothing force, and a stochastic term which tends to roughen the interface. Numerical
simulations along with some theoretical results have confirmed that in the long time $t$
scaling limit, fluctuations in the height of such evolving interfaces scale like $t^{1/3}$
and display non-trivial spatially correlations in the scale $t^{2/3}$
\cite{FNS,KPZ,BSt,ICReview,JeremyReview}. These scales were confirmed experimentally in
studied involving paper wetting, burning fronts, bacterial colonies and liquid crystals
\cite{BSt, TakReview}.

Beyond the KPZ scalings, the universality class is characterized in terms of the long-time
limits of the probability distribution of fluctuations. These depend on the initial data
or geometry. Starting from (i) narrow wedge, or droplet, one sees the $F_{\rm GUE}$
distribution of the Gaussian Unitary Ensemble of random matrix theory while starting from
(ii) flat substrate, the $F_{\rm GOE}$ distribution of the Gaussian Orthogonal Ensemble arises. A recent series of spectacular experiments involving turbulent liquid crystals \cite{TS,TSSS} have been able to not only confirm the predicted scaling laws but also the statistics (skewness and kurtosis) for the distribution of these fluctuations. The multi-point joint distributions of scaled fluctuations are likewise given by the (i) Airy$_2$ \cite{PS} and  (ii) Airy$_1$ \cite{BFPS,BFPr} processes, and \cite{TS,TSSS} could also demonstrate that certain statistic involving the two-point correlation functions agrees with the predictions. A further natural initial geometry is two sided Brownian motion for which one sees, at later time, a new (though correlated) Brownian motion, with a global height shift given by the $F_0$ distribution \cite{BaikRains}. Note that all these spatial processes have $n$-point distributions given by Fredholm determinants.

In this work we consider two questions: 1.  What are the exact statistics and multi-point joint distributions for growth off of more general initial geometries; and 2. Can one predict multi-time statistics and distributions? Our partial answers follow from our investigation into the {\it KPZ renormalization fixed point} which we denote as $\mathfrak{h}$. We describe its Markovian evolution in two complementary ways: 1. Through a variational formulation similar to that of a stochastically forced Burgers equation, but with a new, nontrivial (but unfortunately not very explicit) driving noise, which we call the {\it Airy sheet}, and with the maximization occurring along a network of paths called the {\it polymer fixed point}; and 2. Through a formula for the transition probabilities, derived by employing non-rigorous methods of replica Bethe ansatz for the KPZ equation \cite{D,CDR,ProS2}. These transition probabilities should enable one to compute multi-point statistics for general initial geometries, and using the Markov property, this should enable one to compute multi-time statistics, thus answering both questions. A complete description or characterization of the noise arising in the forced Burgers equation could provide a stochastic analysis approach to show universality of the KPZ fixed point.

Besides the usual issues with the replica method for the KPZ equation (e.g. the moment
problem is not well-posed and consequently one must deal with trying to sum divergent
series), in deriving our transition probability formulas we employ an asymptotic
factorization ansatz on the Bethe wavefunctions. For the narrow wedge initial data, this
ansatz has been shown to  lead in the long time limit to the expected multi-point joint
distributions \cite{ProS2,Dotmultipoint,ISSmultipoint}. However, since we are dealing with
general initial data it is not clear whether this factorization approximation (after
asymptotics) leads to the true transition probability formulas. (Such transition
probability formulas are only known for a few types of initial data, such as flat and two
sided Brownian motion.) As such, our transition probability formulas should be treated as
plausible answer, as they pass several basic tests such as scaling invariance and the
Markov property. They also reproduce the known formulas for one dimensional distributions
for general initial data. In Appendix \ref{appA} we consider the two-point distribution
for flat initial data and produce a formula using our transition probability
formulas. Unfortunately, we have not been able to match this (or to show that this does
not match) with the expected Airy$_1$ process formulas.

Although the connection is not yet understood, our transition probability formula should be accessible via asymptotics of a less explicit determinantal formula derived earlier for the microscopic model TASEP \cite{Sch97, BFPS}. Another possible route to make rigorous our transition probability formula (or disprove it) is through the rigorous replica Bethe ansatz developed in \cite{BCS,BCPS} for $q$-TASEP.

The results of this article should not be treated as mathematically rigorous and rather are intended to provide conjectural descriptions of the KPZ renormalization fixed point. Significant and serious mathematical challenges exist to make rigorous any part of these conjectures.

\subsection{Acknowledgements}
The authors would like to thank H. Spohn, J. Baik, E. Cator, E. Corwin, K. Khanin,
B.Valko, B.Virag, A. Borodin, the special semesters on Random Matrix Theory and its
Applications at MSRI, and Dynamics and Transport in Disordered Media at the Fields
Institute. IC was partially supported by the NSF PIRE grant OISE-07-30136, the grant
DMS-1208998 as well as by Microsoft Research and MIT through the Schramm Memorial
Fellowship, by the Clay Mathematics Institute through the Clay Research Fellowship, by the
Institute Henri Poincare through the Poincare Chair, and by the Packard Foundation through
a Packard Fellowships for Science and Engineering. JQ is supported by the NSERC. DR was
partially supported by NSERC, by a Fields-Ontario Postdoctoral Fellowship, by Fondecyt
Grant 1120309, by Conicyt Basal-CMM, and by Programa Iniciativa Cient\'ifica Milenio grant
number NC130062 through Nucleus Millenium Stochastic Models of Complex and Disordered
Systems.

\section{Variational formulation}

\subsection{The KPZ equation}
In 1986, Kardar-Parisi-Zhang (KPZ) \cite{KPZ} proposed the model equation (which now bears their names)
\begin{equation}
\partial_t h = \tfrac12 (\partial_xh)^2 + \tfrac12 \partial_x^2 h +  \xi,
\label{KPZ}
\end{equation}
from which the universality class takes its name.  The noise $\xi$ is Gaussian space-time
white noise with formal covariance $\langle \xi(t,x) \xi(s,y)\rangle =
\delta(t-s)\delta(y-x)$. It is, in fact, a mathematical challenge to even define this
equation, let along study its long-time scaling behaviors. The physically relevant
notation of solution is provided through the Hopf-Cole transform $Z = e^h$ which
(formally) transforms the KPZ equation into the well-posed stochastic heat equation (SHE)
with multiplicative noise
\begin{equation}\label{SHE}
\partial_tZ = \tfrac12 \partial_x^2 Z  +   \xi Z.
\end{equation}
We will always work with the so-called {\it Hopf-Cole solution} to the KPZ equation which is defined as
$$h(t,x) = \log Z(t,x).$$
The narrow wedge initial conditions correspond to starting $Z$ with a delta function, and the flat initial conditions correspond to starting with $Z\equiv 1$. Correspondingly, in the liquid crystal experiments the laser excites a single point, or a line.

Many discrete growth models have a tunable asymmetry and (\ref{KPZ}) appears as a continuum limit in the diffusive time scale as this parameter is critically tuned close to zero \cite{BG,ACQ,AKQ}. Let us demonstrate this idea via explaining how the KPZ equation scales. For real $b,z$ define the scaled KPZ solution
$$
h_{\e;b,z}(t,x) = \e^{b} h(\e^{-z}t,\e^{-1}x)
$$
Under this scaling,
$$
\partial_t h_{\e;b,z} = \tfrac{1}{2}\e^{2-z} \partial_{x}^2 h_{\e;b,z} + \tfrac{1}{2} \e^{2-z-b} (\partial_x h_{\e;b,z})^2 + \e^{b-z/2+1/2}\xi.
$$
Note that the noise on the right-hand side is not the same for different $\e$, but in terms of its distribution it is. It is natural to consider whether there are there any scalings of the KPZ equation under which it is invariant. If so, then one could hope to scale a given growth process in the same way to arrive at the KPZ equation. However, one checks that there is no way to do this. On the other hand, there are certain weak scalings which fix the KPZ equation. By weak we mean that, simultaneously as we scale time, space and fluctuations, we also put tuning parameters in front of certain terms in the KPZ equation and scale them with $\e$. In other words, we simultaneously scale time, space and fluctuations, as well as the model. Let us consider two weak scalings.

{\bf Weak non-linearity scaling:} Take $b=1/2,z=2$. The first and third terms stay fixed,
but the middle term blows up. Thus, insert a constant $\lambda_{\e}$ in front of the
non-linear term $(\partial_xh)^2$ and set $\lambda_{\e}=\e^{1/2}$. Under this scaling, the
KPZ equation is mapped to itself.

{\bf Weak noise scaling:} Take $b=0,z=2$. Under this scaling, the linear $\partial_x^2 h$
and non-linear $(\partial_x h)^2$ terms stay fixed, but now the noise blows up. So insert
a constant $\beta_{\e}$ in front of the noise term and set $\beta_{\e}=\e^{1/2}$, and again the KPZ equation stays invariant.

One can hope that these rescalings are attractive, in the sense that if one takes models with a parameter (non-linearity or noise) that can be tuned, then these models will all converge to the same limiting object. There are a handful of rigorous mathematical results showing that weak non-linearity scaling can be applied to particle growth processes \cite{BG,ACQ,DemboTsai} and weak noise scaling can be applied to directed polymers.

A third scaling of interest is the following.

\begin{figure}[ht]
\begin{center}
\includegraphics[scale=.95]{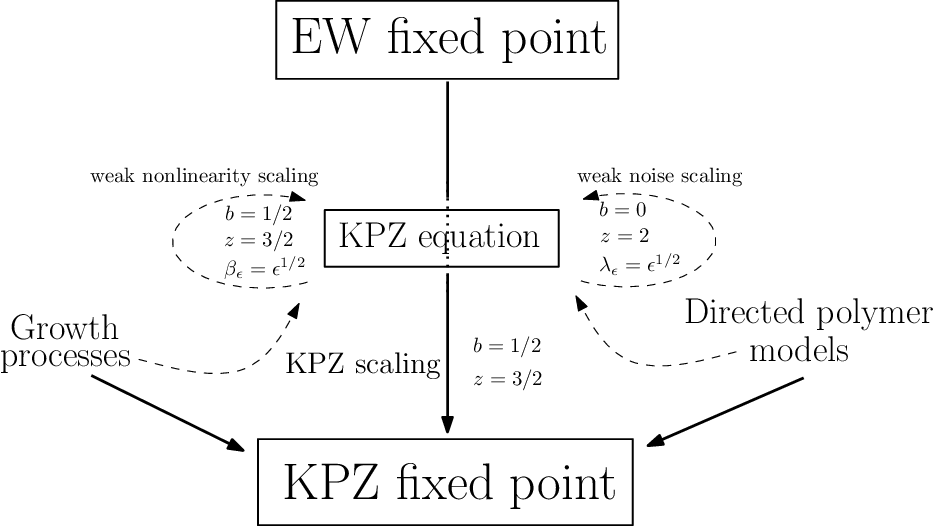}
\end{center}
\caption{Three types of scalings for the KPZ equation. Weak noise and weak non-linearity scaling fix the KPZ equation whereas under KPZ scaling, the KPZ equation should go to the KPZ fixed point. It is believed (and in some cases shown) that this extends to a variety of growth processes and directed polymer models. The KPZ equation represents a heteroclinic orbit connecting the Edwards-Wilkinson (EW) and KPZ fixed point.}\label{KPZequationandfixedpoint}
\end{figure}

{\bf KPZ scaling:} It was predicted by \cite{FNS,KPZ} that under the scaling $b=1/2$ and $z=3/2$ the KPZ equation should have non-trivial limiting behavior. It is this scaling and its limit which is of primary interest in this paper. Figure \ref{KPZequationandfixedpoint} summarizes these scalings and the role of the KPZ equation, KPZ fixed point and EW fixed point.

\subsection{The renormalization operator}

We now fix the KPZ scaling $b=1/2$ and $z=3/2$ and thus set
\begin{equation}\label{zh}
h_\epsilon(t,x)=(R_\e h)(t,x)  =\epsilon^{1/2} h( \epsilon^{-3/2} t, \epsilon^{-1}x).
\end{equation}
Under these changes of variables, $h_\epsilon$ satisfies (\ref{KPZ}) with renormalized coefficients,
\begin{equation*}
\partial_t h_\epsilon = \tfrac12 (\partial_xh_\epsilon)^2 + \epsilon^{1/2}\tfrac12 \partial_x^2 h_\epsilon + \epsilon^{1/4} \xi.
\end{equation*}
It would seem, at first glance, that as $\epsilon\to 0$ all of the coefficients on the
right-hand side go to zero, except that of the non-linearity. Thus, formally one might
expect that $h_0$ satisfies the {\it deterministic} inviscid Burgers equation. However,
this cannot be true since it would imply two false results: that the narrow wedge solution
has {\it deterministic solutions} and that Brownian motion is {\it not} invariant.

The {\it KPZ renormalization fixed point} $\mathfrak{h}$ should be the $\epsilon\to 0$ limit of the (properly centered) process $\bar{h}_\epsilon$. We now provide a description of what this limit should be.


Let $h(u,y;t,x)$ be the solution of (\ref{KPZ}) for times $t>u$ started at time $u$ with a
delta function at $y$, all using the same noise (for different $u,y$).  To center, set
\[\bar{h}(u,y;t,x) ={h}(u,y;t,x) -\tfrac1{24}(t-u) -\log\sqrt{2\pi (t-u)}\] and define $A_{1}$ by
\begin{eqnarray*} &
\bar{h}(u,y;t,x) =  -\tfrac{ (x-y)^2}{ 2(t-u)}  + A_{1}(u,y;t,x). &
\end{eqnarray*}
After the rescaling (\ref{zh}),
\begin{equation*}
R_\e\bar{h}(u,y;t,x) =  -\tfrac{ (x-y)^2}{ 2(t-u)}  + A_{\epsilon}(u,y;t,x)
\end{equation*}
where $A_{\epsilon}=R_\e A_{1}$.
As $\epsilon\to 0$, $A_{\epsilon}(u,y;t,x)$ should converge to a four-parameter field which we henceforth call the {\it space-time Airy sheet} $\mathcal{A}(u,y;t,x)$. In each spatial variable it should be an $\textrm{Airy}_2$ process \cite{PS} and it should enjoy several nice properties:
 \begin{enumerate}
 \item  {\it Independent increments.}  $\mathcal{A}(u,y;t,x)$ is independent of $\mathcal{A}(u',y;t',x)$ if $(u,t)\cap(u',t')=\emptyset$;
 \item {\it Space and time stationarity.}
$
\mathcal{A}(u,y;t,x) \stackrel{\rm dist}{=} \mathcal{A}(u+h,y;t+h,x)
 \stackrel{\rm dist}{=} \mathcal{A}(u,y+z;t,x+z)
$;
\item {\it Scaling.} $
\mathcal{A}(0,y;t,x) \stackrel{\rm dist}{=} t^{1/3} \mathcal{A}(0,t^{-2/3}y;1,t^{-2/3}x)$;
\item {\it Semi-group property.} For $u<s<t$,
\vskip-0.25in
\begin{equation*}
\mathcal{A}(u,y;t,x) = \sup_{z\in \R} \big\{\tfrac{(x-y)^2}{2(t-u)} -\tfrac{(z-y)^2}{2(s-u)}-\tfrac{(x-z)^2}{2(t-s)} + \mathcal{A}(u,y;s,z)+\mathcal{A}(s,z;t,x) \big\}.
\end{equation*}
\end{enumerate}

%
%
%
%
%

Using $\mathcal{A}(u,y;t,x)$ we construct our conjectural description of the KPZ fixed
point $\mathfrak{h}(t,x)$.  By the Hopf-Cole transformation and the linearity of the
SHE, the centered solution of (\ref{KPZ}), $\bar{h}(t,x) = h(t,x) -
\tfrac{t}{24} - \log\sqrt{2\pi t}$, with initial data $h^0$, can be written after
rescaling as
\vskip-0.2in
\begin{equation*}
R_{\e} \bar{h}(t,x) = \e^{1/2}\ln \int \exp\Big( \e^{-1/2} \big\{ -\tfrac{(x-y)^2}{2t} + A_\e(0,y;t,x) +R_\e h^0(y)\big\} \Big) dy.
\end{equation*}
If we choose initial data $h^0_\e$ so that $R_\e h^0_{\e}$ converges to some (possibly random) function $f$ as $\e\to 0$, we can use Laplace's method to evaluate
$
\mathfrak{h}(t,x)=\lim_{\e\to 0}R_\e\bar{h}(t,x) = T_{0,t} f(x)
$
where
\begin{equation} \label{Toper}
T_{u,t}f(x):= \sup_{y\in \R}\left\{  - \tfrac{ (x-y)^2}{ 2(t-u)}   + \mathcal{A}(u,y;t,x) + f(y) \right\}.
\end{equation}
The operators $T_{u,t}$, $0<u<t$ form a semi-group, i.e. $T_{u,t}=T_{u,s}T_{s,t}$, which
is stationary with independent increments, and such that
\[T_{0,t} \stackrel{\rm dist}{=} (R_{t^{-2/3}})^{-1}T_{0,1}R_{t^{-2/3}}.\]
Additionally, if $\alpha\in\rr$ then
\begin{equation}
T_{0,t}(\alpha f)(x)\stackrel{\rm dist}{=}\alpha T_{\alpha^{-3}t}(\alpha^{-2}x).\label{eq:alpha}
\end{equation}

By the Markov property, the joint distribution of the marginal spatial process of $\mathfrak{h}$ (for initial data $f$) at a set of times $t_1< t_2<\cdots < t_n$ should be given by
\begin{equation*}
(\mathfrak{h}(t_1),\ldots, \mathfrak{h}(t_n))\stackrel{\rm dist}{=} (T_{0,t_1}f,   \ldots, T_{t_{n-1},t_n}\cdots T_{0,t_1}f).
\end{equation*}
The process of randomly evolving functions can be thought of as a high dimensional analogue of Brownian motion (with function-valued state space), and the $T_{t_{i},t_{i+1}}$ as analogous to the independent increments.

The solution $h(u,y;t,x)$ of (\ref{KPZ}) (started at time $u$ with a delta function at
$y$) corresponds to the free energy of a directed random polymer $x(s)$, $u<s<t$ starting
at $y$ and ending at $x$, with quenched random energy (see \cite{AKQcontpolym} for a
rigorous construction of this measure)
\begin{equation}
\int_u^t \{|\dot x(s)|^2 -\xi(s,x(s)) \}\,ds. \label{hetch}
\end{equation}
Under the rescaling (\ref{zh}) this probability measure on paths should converge to the {\it polymer fixed point}; a continuous path $\pi_{u,y;t,x}(s)$, $u\le s\le t$ from $y$ to $x$ which at discrete times $u=s_0<\cdots s_{m-1}<t$ is given by the argmax over $x_0,\ldots, x_{m-1}$ of
\begin{equation}\label{polymerfixedpt}
(T_{u,s_1}\delta_y)(x_1) + (T_{s_1,s_2} \delta_{x_1})(x_2) +\cdots +
(T_{s_{m-1},t}\delta_{x_{m-1}})(x).
\end{equation}
This is the analogue in the present context of the minimization of the action, and the
polymer fixed point paths are analogous to characteristics in the randomly forced Burger's
equation. One might hope to take the analogy farther and find a limit of the
renormalizations of (\ref{hetch}), and  minimize it to find that path
$\pi_{u,y;t,x}$. However, the limit of the energy (\ref{hetch}) does not appear to exist, so one has to be satisfied
with a limit of the  path measures themselves.  The path $\pi_{0,y;t,x}$ should be  H\"{o}lder continuous with exponent $1/3-$, as compared to Brownian motion where the  H\"{o}lder exponent is $1/2-$. As the mesh of times is made finer, a limit $\mathcal{E}(\pi_{0,y;t,x})$
of (\ref{polymerfixedpt}) should exist, and through it we can write the time evolution of the KPZ fixed point in terms of the polymer fixed point through the analogue of the Lax-Oleinik variational formula,
\begin{equation*}
T_{u,t}f(x) = \sup_{y\in R} \{ \mathcal{E}(\pi_{u,y;t,x})+ f(\pi_{u,y;x,t}(u))\}.
\end{equation*}

\begin{figure}
\psfrag{space0}[c]{$\e^{-1}$}
\psfrag{space3}[c]{$\e^{-1}x$}
\psfrag{time1}[c]{$\e^{-3/2}t_1$}
\psfrag{time2}[c]{$\e^{-3/2}t_2$}
\psfrag{time3}[c]{$\e^{-3/2}t_3$}
\psfrag{fluc}[c]{$\e^{-1/2}$}
\begin{center}
\includegraphics[scale=.28]{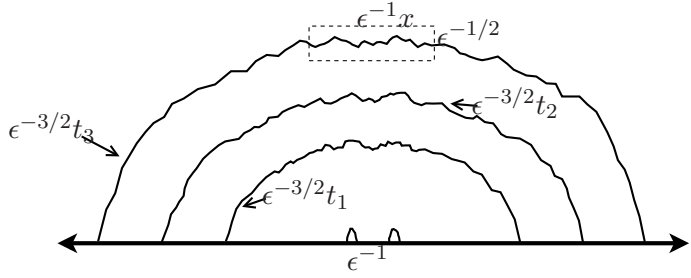}
\end{center}
\caption{Polynuclear growth with two nucleations distance $\e^{-1}$ apart observed at times $\e^{-3/2}t_i$ ($i=1,2,3$) in an $\e^{-1}$ spatial scale and $\e^{-1/2}$ fluctuation scale.}\label{fig1}
\end{figure}

The KPZ fixed point, space-time Airy sheet, and polymer fixed point should be universal
and arise in random polymers, last passage percolation and growth models -- anything in
the KPZ universality class. Just as for (\ref{KPZ}), for some models at the microscopic
scale, approximate versions of the variational problem (\ref{Toper}) hold, becoming exact
as $\epsilon\to 0$. For example, consider the PNG model \cite{PS}  with a finite
collection of nucleations spaced order $\e^{-1}$ apart (see Figure \ref{fig1}). At time $\e^{-3/2}t$ we look at $\e^{-1/2}$ scaled fluctuations in spatial locations $\e^{-1}x$. As $\e$ goes to zero, these fluctuations (after proper centering) should converge to $\mathfrak{h}$ where the initial data $f$ is $-\infty$ except at the nucleation points, where it is zero. By introducing additional nucleations at times on the order of $\e^{-1/2}$ and spatial locations order $\e^{-1}$ apart, it is possible to modulate the value of $f$ at these non $-\infty$ points. Taking the number of nucleation points large allows one to recover any $f$. The experiment of \cite{TS,TSSS} is well described by the KPZ fixed point with a single nucleation. Future experiments could probe the effect of additional nucleations. Using statistics to differentiate between types of initial data given finite time observations is a driving force for the development of the following exact formulas which provide theoretical predictions.

\section{Transition probability formulation}

We start with a simple version of our proposed formulas which are the results of computations described in Section \ref{RBA}.

Given ${c}_i, y_i\in \R$, $i=1,\ldots,\nu$, and $s_j,x_j\in \R$, $j=1,\ldots,\mu$, let
\[\mathfrak{h}(t,x)=(T_t f)(x)\quad\text{with}\quad f(y_i)=-\big(\tfrac{t}{2}\big)^{1/3}
[c_i -(y_i-x_1)^2]\quad\text{and}\quad f=-\infty~\text{otherwise}.\]
Then the computations of Section \ref{RBA} suggest that
\begin{equation}
  \pp\big(\mathfrak{h}(t,x_j)\le\big(\tfrac{t}{2}\big)^{1/3}[s_j -(x_j-x_1)^2], ~j=1,\ldots,\mu\big) = \det(I - \bar{L})\label{eq:trpr}
\end{equation}
where $\bar{L}$ is the operator  with kernel
 \begin{multline}\label{discrtp}
 \bar{L}(z,z') =  \int_{A({\bf{s},\bf{c}})} dv_1\dotsc dv_{\nu+1} \, du_1\dotsc du_{\mu}\, e^{(y_1-x_1)H}\K|v_1\rangle \langle v_1| e^{(y_2-y_1)H}|v_2\rangle \langle v_\nu | e^{(x_1-y_\nu)H}|v_{\nu+1}\rangle\\
  \times\delta(u_1 - v_{\nu+1}) \langle u_1|e^{ ( x_1 -x_2) H}|u_2\rangle \langle u_{\mu-1}|e^{ (x_{\mu- 1} - x_{\mu}) H}|u_{\mu}\rangle\langle u_{\mu}|e^{(x_{\mu} - x_1) H}\K
\end{multline}
acting on $L^2(\R)$, where $H=-\frac{d^2}{dx^2} +x$ is the
Airy operator, $\K$ is the ${\rm Airy}_2$ operator which is the spectral projection of $H$ onto its negative eigenvalues, and
\begin{equation}\label{ayeaye}
A({\bf{s},\bf{c}}) =\big\{ ({\bf{u},\bf{v}}): \max_{i=1,\ldots,\mu}\{u_i-s_i\}+ \max_{j=1,\ldots,\nu}\{v_j -c_j\} \ge v_{\nu+1}\big\}.
\end{equation}
We  assume here that $x_1<x_2<\cdots <x_\mu$ and $y_1>\cdots>y_\nu$.  The formulas give a
consistent family of finite dimensional distributions.  Specializing to $\mu=1$ or $\nu=1$
one checks that the resulting process (in $x_1$ and $y_1$) is the Airy$_2$ process in each
variable. Note that, in view of \eqref{eq:alpha}, the left-hand side of \eqref{eq:trpr}
can be rewritten as $\pp\big(\mathfrak{h}(2,x_j)\le[s_j -(x_j-x_1)^2],
~j=1,\ldots,\mu\big)$ where $\mathfrak{h}(2,x)=T_2\tilde f(x)$ with $\tilde f(y_j)=-[c_j
-(2/t)^{4/3}y_j^2]$ and  $f=-\infty$ otherwise.

From the above the transition probabilities of $T_t=T_{0,t}$ can be obtained by approximation.  Define the operator $Y_{[a,b]}^{g}$ via its action $Y_{[a,b]}^{g}f(y) =u(b,y)$ where $u(t,x)$ solves the boundary value problem\begin{equation}\label{why}\left\{\begin{array}{ll}
\partial_t u = -H u &\mbox{ $a<t<b$} \\
u(t,x)=0&\mbox{$x\geq g(t)$}\\
\end{array}
\right.
\end{equation}
with initial data $u(a,x) = f(x)$.  Let $c(\cdot)$ and $s(\cdot)$ be (nice) functions such
that $s(\cdot)$ has finite  support in $[L_s,R_s]$ and $c(\cdot)$ has finite support in
$[L_c,R_c]$ (here by finite support we mean that outside that interval the function is $+\infty$). Then
\begin{equation}
\pp\big({T}_t f (x)\le (\tfrac{t}{2})^{1/3}[s(x)-(x-L_s)^2], x\in [L_s,R_s]\big) = \det(I-\K+\hat{L})\label{eq:fpformula}
\end{equation}
where $f(y) =-(\frac{t}{2})^{1/3}\left( c(y) - (y-L_s)^2\right)$ and
\begin{equation}\hat{L}(z,z') =  \iint\! dm \,du\,\Theta_{1,m,u}(u,z)\frac{d}{dm} \Theta_{2,-m}(z',u),\label{p2}\end{equation}
\begin{equation*}\Theta_{1,m,u_1}(u_2,z) = \left(Y_{[L_s,R_s]}^{s(\cdot)+u_1+m} e^{(R_s-L_s) H}\K\right)(u_2,z),\label{p3}\end{equation*}
\begin{equation}\Theta_{2,m'}(z',v_{\nu+1})=\left(\K e^{(R_c-L_s)H}Y_{[L_c\,R_c]}^{\hat{c}(\cdot)+m'} e^{(L_s-L_c) H}\right)(z',v_{\nu+1}),\label{p4}\end{equation}
where $\hat{c}(y)=c(L_c+R_c-y)$.


\subsection{Replica Bethe ansatz}\label{RBA}
In this section we follow the methods of \cite{D,CDR} as further developed  in \cite{ProS1,ProS2}.
Let ${Z}(y,x;t)$  be the solution of (\ref{SHE}) at $x$ at time $t$ with initial data $\delta_{y}$ at time $0$.
%
Given $s_k, x_k$, $k=1,\ldots,\mu$ and $c_m, y_m$, $m=1,\ldots,\nu$ consider $\sum_{k=1}^\mu e^{-s_k} {Z}(t,x_k)$
where ${Z}(t,x)$ solves the SHE with initial data $\sum_{m=1}^{\nu} e^{-c_m}\delta_{y_m}$.
From the linearity of the SHE this can be written as $\sum_{k=1}^{\mu}\sum_{m=1}^{\nu}e^{-s_k-c_m} {Z}(y_m,x_k;t)$.
Following \cite{D,CDR} we compute the generating function
\begin{equation*}
G(s,x;c,y) = \EE\bigg[\!\exp\!\Big(\!-e^{\tfrac{t}{24}} \big\{\sum_{k=1}^{\mu}\sum_{m=1}^{\nu}e^{-s_k-c_m} {Z}(y_m,x_k;t)\big\}\Big)\bigg].
\end{equation*}
By studying asymptotics under KPZ scaling we obtain the formula for the transition probabilities.

Expanding the generating function exponential we write this using replicas as
\begin{equation}\label{eq:G}
G(s,x;c,y) = 1 + \sum_{N=1}^{\infty} \frac{(-1)^N e^{tN/24}}{N!} \langle I^N| e^{-H_N t}|F^N\rangle
\end{equation}
where the $N$-particle wave functions of the delta-Bose gas
\begin{equation}{\rm H}_N= -\frac{1}{2}\sum_{i=1}^{N} \partial_{x_i}^2 - \frac{1}{2}
  \sum_{i\neq j=1}^{N} \delta(x_i-x_j)\end{equation} are $\langle I^N|= \sum_{m_1,\ldots
  m_N=1}^{\nu} e^{-\sum_jc_{m_j}} \langle y_{m_1}\cdots y_{m_N}|$ and $|F^N\rangle=
\sum_{k_1,\ldots,k_N=1}^{\mu} e^{-\sum_js_{k_j}} |x_{k_1}\cdots
x_{k_N}\rangle$.  The wave functions are symmetric, so the propagator is only needed on
the symmetric subspace. Thus we may employ the eigenfunction expansion of ${\rm H}_N$
(e.g. \cite{D,CDR}),
\begin{equation}\label{tense}
\langle I^N| e^{-H_N t}|F^N\rangle = \sum_r e^{-t E_r} \langle F^N|\psi_r\rangle \langle \psi_r|I^N\rangle.
\end{equation}
The eigenfunctions are given by
\begin{equation}
\psi_{\bf q, n}^{(L)} =\sum'_{p\in{\mathcal P}} A_p({\bf y})
\exp\Big\{ i \sum_{\alpha=1}^L q_\alpha
\sum_{c\in\Omega_{\alpha(p)}} y_c  - \frac14\sum_{\alpha=1}^L
\sum_{c,c'\in\Omega_{\alpha(p)}}|y_c-y_{c'}| \Big\}.
\end{equation}
Here $1\le L\le N$ is the number of clusters of particles, $n_\alpha$ is the number of
particles in the $\alpha$-th cluster, and $N=\sum_{\alpha=1}^L n_\alpha$ is the total
number of particles.  $\sum'_{p\in{\mathcal P}}$ denotes the sum over those permutations of $N$ elements permuting particles between different clusters.  ${\bf q} = (q_1,\ldots,q_M)$ are the momenta, and ${\bf y} = (y_1,\ldots,y_N)$.  See \cite{D} for the
definitions of the coefficients $A_p({\bf y}) $ and the
partition $\Omega_\alpha(p)$, $\alpha=1,\ldots, L$ of $\{1,\ldots,N\}$.

Expanding the tensor products given above, and still writing $r$ to index the
eigenfunctions for simplicity, we get
\begin{multline*}
G(s,x;c,y) = 1 +\sum_{N=1}^{\infty} \frac{(-1)^N e^{tN/24}}{N!} \sum_{r}e^{-tE_r}|\psi_r({\bf 0})|^2  \sum_{k_1,\ldots, k_N=1}^{\mu} e^{-(s_{k_1}+\cdots+ s_{k_N})} \frac{\psi_r(x_{k_1},\ldots, x_{k_N})}{\psi_r({\bf 0})}
\\\hskip-.1in\times \sum_{m_1,\ldots, m_N=1}^{\nu} e^{-(c_{m_1}+\cdots+ c_{m_N})} \left[\frac{\psi_r(y_{m_1},\ldots, y_{m_N})}{\psi_r({\bf 0})}\right]^*.
\end{multline*}
Using the factorization approximation of \cite{ProS2} with the identity $e^{au+bv+cuv}=e^{c\partial_{a}\partial_b}e^{au+bv}$,
\begin{equation*}
\sum_{k_1,\ldots, k_N=1}^{\mu} e^{-(s_{k_1}+\cdots+ s_{k_N})} \frac{\psi_r(x_{k_1},\ldots, x_{k_N})}{\psi_r({\bf 0})} \approx \prod_{\alpha=1}^{M} e^{-\tfrac{1}{4} \sum_{a,b=1}^{\mu} |x_a-x_b|\partial_{s_a}\partial_{s_b}} \left(\sum_{k=1}^{\mu}e^{-s_k+i q_{\alpha}x_k}\right)^{n_{\alpha}}.
\end{equation*}
This approximation is believed to hold asymptotically in $t$, the evidence being the recovery of the Airy$_2$ process in the limit when $\nu=1$ \cite{ProS2}.

We now follow many of the manipulations made in \cite{ProS2} in the case when $\nu=1$.
Plugging in the above factorization we have an approximate generating function,
\begin{align*}
G^{\#}(s,x;c,y) =  1 + \sum_{N=1}^{\infty} \frac{(-1)^N e^{\frac1{24}tN}}{N!} \sum_{r}e^{-tE_r}|\psi_r({\bf 0})|^2 & \prod_{\alpha=1}^{L} e^{-\frac{1}{4} \sum_{a,b=1}^{\mu} |x_a-x_b|\partial_{s_a}\partial_{s_b}} \left(\sum_{k=1}^{\mu}e^{-s_k+i q_{\alpha}x_k}\right)^{n_{\alpha}} \\
&\hskip-.1in \times \prod_{\alpha=1}^{L} e^{-\frac{1}{4} \sum_{a,b=1}^{\nu} |y_a-y_b|\partial_{c_a}\partial_{c_b}} \left(\sum_{l=1}^{\nu}e^{-c_l-i q_{\alpha}y_l}\right)^{n_{\alpha}}.
\end{align*}
Now recall \cite{D} that the eigenenergies
$E_r = \frac{1}{2}\sum_{j=1}^{M} n_j q_j^2 - \frac{1}{24} \sum_{j=1}^{M} (n_j^3-n_j)$, that
\begin{equation*} |\psi_r({\bf 0})|^2 = N! \det\left(\frac{1}{\tfrac{1}{2}(n_j+n_k)+i(q_j-q_k)}\right)_{j,k=1,\ldots,M},\end{equation*}
and that the normalized sum over eigenstates is given by
$$\sum_r= \sum_{L=1}^{\infty} \frac{1}{L!} \prod_{j=1}^{L} \left(\int_{-\infty}^{\infty} \frac{dq_j}{2\pi} \sum_{n_j=1}^{\infty}\right){\bf 1}_{N=\sum_{j}n_j}.$$
Thus we can write
\begin{multline*}
  G^{\#}=1+ \sum_{L=1}^{\infty} \frac{1}{L!}\left[ \prod_{j=1}^{L}\int_{-\infty}^{\infty}
    \frac{dq_j}{2\pi} \sum_{n_j=1}^{\infty}
    e^{\frac{1}{24}tn_j^3-J}\left\{-e^{-\frac{1}{2}tq_j^2}\left(\sum_{a=1}^{\mu} e^{ix_a
          q_j-s_a}\right)\left(\sum_{a=1}^{\nu} e^{-iy_a
          q_j-c_a}\right)\right\}^{n_j}\right]
  \\\times\det\left(\frac{1}{\tfrac{1}{2}(n_j+n_k)+i(q_j-q_k)}\right)
\end{multline*}
where
\begin{equation*}
J=\tfrac{1}{4}\textstyle\sum\nolimits_{a,b=1}^{\mu}|x_a-x_b|\partial_{s_a}\partial_{s_b}
+\tfrac{1}{4}\sum\nolimits_{a,b=1}^{\nu}|y_a-y_b|\partial_{c_a}\partial_{c_b}.
\end{equation*}
At this point one is forced to adopt a choice $e^{tm^3/3}= \int\! du\,{\rm Ai}(u)
e^{umt^{1/3}}$ of analytic continuation to complex $m$.  Then one observes, following
\cite{D}, that $G^{\#}$ can be written as a Fredholm determinant $G^{\#} = \det(1+R)$
where $R$ has kernel $R(q,m;q',m') $ given by
$$
\frac{1}{2\pi(\tfrac{1}{2}(m+m') +i
  (q-q'))}e^{\tfrac{tm^3}{24}-J}\left\{-e^{-\tfrac{tq^2}{2}}
  {\textstyle\left(\sum_{a=1}^{\mu} e^{ix_a q-s_a}\right)\left(\sum_{a=1}^{\nu} e^{-iy_a q-c_a}\right)}\right\}^{m}.$$
Define for vectors ${\bf s}=(s_1,\ldots, s_\mu)$ and ${\bf c}=(c_1,\ldots, c_{\nu})$ the function
\begin{equation*}\Phi({\bf s};{\bf c})= \frac{ (e^{s_1}+\cdots+ e^{s_{\mu}})(e^{c_1}+\cdots+ e^{c_{\nu}})}{ 1+  (e^{s_1}+\cdots+ e^{s_{\mu}})(e^{c_1}+\cdots+ e^{c_{\nu}})}.\end{equation*}
Following \cite{ProS2} we obtain that
$G^{\#} = \det(1-\tilde{N})$ with
\begin{equation*}\tilde{N}(z,z') = {\bf 1}_{z,z'>0}e^{-\hat{J}}\int du \Ai(u+z)\Ai(u+z') \Phi(\alpha {\bf u -s}; -{\bf c})\end{equation*}
where ${\bf u} = (u, \ldots, u)$ and where, for $\alpha = (t/2)^{1/3}$,
\begin{equation*}
\textstyle\hat{J} = J+ (2\alpha)^{-1}(\partial_z-\partial_{z'}) \left(\sum_{a=1}^{\mu} x_a\partial_{s_a}-\sum_{b=1}^{\nu} y_b\partial_{c_b}\right).
\end{equation*}

We can follow the method of \cite{ProS2} equation (4.21) to replace $\partial_z$ by either $- \partial_{z'} +\alpha\sum_{b=1}^\mu\partial_{s_b}$ or $ -\partial_{z'} +\alpha\sum_{b=1}^\nu\partial_{c_b}$ because $\sum_{b=1}^\mu\partial_{s_b}$ and $\sum_{b=1}^\nu\partial_{c_b}$ have the same action on $\Phi$. Hence we get
\begin{equation}
  \tilde{N}(z,z') = \tau_{ {\bf x}^2/2t,  {\bf y}^2/2t } {\bf 1}_{z,z'>0}e^{J_{\bf y}}e^{J_{\bf x}}\int du
  \Ai(u+z)\!\Ai(u+z') \Phi(\alpha {\bf u -s}; -{\bf c})
\end{equation}
where \begin{equation*}
J_{\bf x} ={\textstyle -\sum_{a>b}^{\mu}x_a\partial_{s_a}\partial_{s_b}-\tfrac{1}{2}\sum_{a=1}^{\mu} x_a\partial_{s_a}^2  -\tfrac{1}{4\alpha^3}\sum_{a=1}^{\mu} x_a^2 \partial_{s_a} +\tfrac{\partial_{z'}}{\alpha} \sum_{a=1}^{\mu} x_a\partial_{s_a}},
\end{equation*}\\[-5pt]
$J_{\bf y}$ is the analogue of $J_{\bf x}$ with ${\bf s},{\bf x}$ replaced by ${\bf c},-{\bf y}$,
and
\begin{equation*}
\tau_{ {\bf x}^2/2t,  {\bf y}^2/2t }= \exp\left\{{\textstyle\frac{1}{4\alpha^3} \sum_{a=1}^{\mu} x_a^2 \partial_{s_a} +\frac{1}{4\alpha^3} \sum_{a=1}^{\nu} y_a^2 \partial_{c_a} }\right\}
\end{equation*} is the parabolic shift (recall that $f(s+d)=\exp(d\partial_s)f(s)$).  From
now on we move into the frame with parabolic shifts removed (which accounts for the
parabolas in the transition probability formula \eqref{eq:trpr}, the shift by $x_1$ will
be explained later), and call the associated operator $\tilde
L$ (instead of $\tilde N$). Thus, in view of \eqref{eq:G}, we are computing
\begin{equation}
\EE\!\left(\exp\!\left[-e^{\sum_{k=1}^K[\frac{t}{24}+h(t,x_k)-(s_k -x_k^2/2t)]}\right]\right) \approx \det(I - \tilde{L})\label{eq:EEkpz}
\end{equation}
with $h(t,x)=\log(Z(t,x))$ the Hopf-Cole solution to the KPZ equation with initial
condition given by $Z(0,\cdot)=\sum_{m=1}^Me^{-c_m+y_m^2/2t}\delta_{y_m}$.

Observe (as in \cite{ProS2}) that one may write (setting $x_{\mu+1}=y_{\nu+1}=0$)
$
e^{J_{\bf x}} = e^{\tfrac{x_1}{2\alpha^2}z'}A^{\bf x}_1\cdots A^{\bf x}_\mu,
$ and $
e^{J_{\bf y}} = e^{\tfrac{-y_1}{2\alpha^2}z'}A^{\bf y}_1\cdots A^{\bf y}_\nu,
$
where
\begin{equation*}
A^{\bf x}_l=\exp\Big\{{\textstyle-\frac{x_\ell-x_{\ell+1}}{2}(\sum_{a=1}^{\ell}\partial_{s_{a}})^2-\frac{(x_{\ell}-x_{\ell+1})^2}{4\alpha^3}\sum_{a=1}^{\ell}\partial_{s_{a}} -\frac{x_{\ell}-x_{\ell+1}}{2\alpha^2}z'-\frac{x_\ell-x_{\ell+1}}{\alpha}\partial_{z'}(\sum_{a=1}^{\ell}\partial_{s_{a}})}\Big\}
\end{equation*}
and $A^{\bf y}_l$ is the analogous operator with ${\bf s},{\bf x}$ replaced by ${\bf c},-{\bf y}$. By the Baker-Campbell-Hausdorff formula,
\begin{equation}
A^{\bf y}_1\cdots A^{\bf y}_\nu e^{\tfrac{x_1}{2\alpha^2}z'}=
e^{\tfrac{x_1}{2\alpha^3} \sum_{\ell=1}^\nu y_\ell\partial_{c_\ell}  + \tfrac{x_1}{2\alpha^2}z'}A^{\bf y}_1\cdots A^{\bf y}_\nu.
\end{equation}
This gives
for our kernel
\begin{equation} \label{kern}
   {\bf 1}_{z,z'>0} e^{\tfrac{x_1}{2\alpha^3} \sum_{\ell=1}^\nu y_\ell\partial_{c_\ell}  + \tfrac{x_1-y_1}{2\alpha^2}z'}
\int\! du \,A^{\bf y}_1\cdots A^{\bf y}_\nu A^{\bf x}_1\cdots A^{\bf x}_\mu\Phi(\alpha
{\bf u -s},  -{\bf c})\Ai(u+z)\Ai(u+z').
\end{equation}

Introducing an auxiliary variable $u_{\mu+1}$, the integral in (\ref{kern}) becomes
\begin{equation}\label{introd}
\int du_{\mu+1}\,du\,\delta(u_{\mu+1} - u) A^{\bf y}_1\cdots A^{\bf y}_\nu A^{\bf
  x}_1\cdots A^{\bf x}_\mu \Phi(\alpha {\bf u -s},  -{\bf
  c})\Ai(u_{\mu+1}+z)\Ai(u+z').
\end{equation}
Now observe the identity
\begin{equation*}
A^{\bf x}_\ell\Phi(\alpha {\bf u-s}, \star)\Ai(\star)\Ai(u+z') = e^{\frac{x_\ell-x_{\ell+1}}{2\alpha^2}H_u}\Phi(\alpha {\bf u-s}, \star)\Ai(\star)\Ai(u+z')
\end{equation*}
where $\star$ represents that all subsequent terms do not depend on $u$, and where $H_u$
is the Airy operator acting in $u$, $H_u = -\partial_u^2 + u$. Likewise,
\begin{equation*}
  A^{\bf y}_\ell\Phi(\star,\alpha {\bf u-c})\Ai(\star)\Ai(u+z') =
  e^{\frac{y_{\ell+1}-y_{\ell}}{2\alpha^2}H_u}\Phi(\star,\alpha {\bf u-c})\Ai(\star)\Ai(u+z').
\end{equation*}

From the first identity we can replace the operator $A^{\bf x}_{\mu}$ by
$e^{\frac{x_\mu-x_{\mu+1}}{2\alpha^2}H_u}$ applied to  $\Phi$ times the Airy function
terms only. Introducing
an additional variable $u_{\mu}$ and a delta function $\delta(u_{\mu}-u)$ we get
\begin{equation*}
\int\!du_{\mu+1}\,du\,\delta(u_\mu-u)\delta(u_{\mu+1}-u_\mu) A^{\bf y}_1\cdots A^{\bf y}_\nu A^{\bf
  x}_1\cdots A^{\bf x}_{\mu}e^{\frac{x_\mu-x_{\mu+1}}{2\alpha^2}H_u}\big[\Phi(\alpha {\bf
  u -s},  -{\bf c})
\!\Ai(u_1+z)\!\Ai(u+z')\big].
\end{equation*}
Then we can integrate by parts so as to move the action onto only the delta function, and
writing $A\delta(v-u)=A(u,v)$ for an operator $A$ acting on $u$, (\ref{introd}) becomes
\begin{multline*}
  \int\!du_{\mu}\,du_{\mu+1}\,du\,
  \delta(u_{\mu}-u)e^{\frac{x_\mu-x_{\mu+1}}{2\alpha^2}H}(u_{\mu},u_{\mu+1})\\
  \times A^{\bf y}_1\cdots A^{\bf y}_\nu A^{\bf
    x}_1\cdots A^{\bf x}_{\mu-1} \Phi\big(\alpha u -s_1,\ldots,\alpha
  u-s_{\mu-1}, \alpha u_{\mu} -s_\mu, -{\bf c}\big)\!\Ai(u_{\mu+1}+z)\!\Ai(u+z').
\end{multline*}
By introducing this additional delta function we have been able to replace the variable
$u$ in the term $\alpha u -s_{\mu}$ by $\alpha u_{\mu}-s_{\mu}$ and likewise
$e^{\frac{x_\mu-x_{\mu+1}}{2\alpha^2}H}(u,u_{\mu+1})$ by
$e^{\frac{x_\mu-x_{\mu+1}}{2\alpha^2}H}(u_{\mu},u_{\mu+1})$.  Iterating this procedure
$\mu-1$ times we obtain
\begin{multline*}
  \int du_1\cdots du_{\mu+1}\,du\, \delta(u_1-u)e^{\frac{x_1-x_2}{2\alpha^2}
    H}(u_1,u_2)\cdots e^{\frac{x_\mu-x_{\mu+1}}{2\alpha^2} H}(u_\mu,u_{\mu+1})\\
  \times  A^{\bf y}_1\cdots A^{\bf y}_{\nu}\Phi\big(\alpha u_1 -s_1,\ldots \alpha u_{\mu}
  -s_\mu, -{\bf c}\big)  \Ai(u_{\mu+1}+z)\Ai(u+z').
\end{multline*}
So far the manipulations have followed exactly those of \cite{ProS2}.  Now in order to
apply a similar procedure for the $A^{\bf y}_l$ operators, apply the change variables
$u_i\mapsto u_i+u$ for $i=1,\ldots, \mu$ (but not $\mu+1$), which yields
\begin{multline*}
\int\!du_1\cdots du_{\mu+1}du\, \delta(u_1)
  \langle u_1|e^{ \frac{ x_1-x_2}{2\alpha^2} H}| u_{2}\rangle \cdots  \langle u_\mu| e^{ \frac{ x_\mu-x_{\mu+1}}{2\alpha^2} H}|u_{\mu+1}-u\rangle \\
  \times   A^{\bf y}_1\cdots A^{\bf y}_{\nu}\Phi\big(\alpha u_1 -s_1,\ldots \alpha u_{\mu} -s_\mu, \alpha{\bf u- c}\big)\! \Ai(u_{\mu+1}+z)\!\Ai(u+z').
  \end{multline*}
The key point is that we now have  $\alpha u -c_i$ in the second set of slots of $\Phi$. We proceed now for the the $A^{\bf y}_{\ell}$ just as we did for the $A^{\bf x}_{\ell}$. Introduce a new variable $v_{\nu+1}$ and a delta function $\delta(v_{\nu+1}-u)$, use the formula above to replace $A^{\bf y}_{\nu}$ by $\exp\{\frac{y_{\nu+1}-y_{\nu}}{2\alpha^2}H_u\}$ applied to the product of the $\Phi$ and $\Ai$ functions, integrate by parts and finally introduce yet another new variable $v_{\nu}$.
Doing this once yields
\begin{multline*}
\int\!dv_{\nu}\, dv_{\nu+1} \,du_1\cdots du_{\mu+1}\,du\, \delta(u_1) \langle u_1|e^{ \frac{ x_1-x_2}{2\alpha^2} H}| u_{2}\rangle \cdots  \langle u_\mu| e^{ \frac{ x_\mu-x_{\mu+1}}{2\alpha^2} H}|u_{\mu+1}-v_{\nu+1}\rangle \\ \times \delta(v_{\nu}-u) (e^{\frac{y_{\nu+1}-y_{\nu}}{2\alpha^2}H}\delta( v_{\nu+1}-v_\nu) )  A^{\bf y}_1\cdots A^{\bf y}_{\nu-1}\\ \times\Phi\big(\alpha u_1 -s_1,\ldots \alpha u_{\mu} -s_\mu, \alpha u-c_1,\ldots,\alpha u -c_{\nu-1}, \alpha v_{\nu}-c_\nu\big) \!\Ai(u_{\mu+1}+z)\!\Ai(u+z').
\end{multline*}
We may iterate this $\nu-1$ more times to get
\begin{multline*}
\int\! dv_{\nu}\, dv_{\nu+1}\, du_1\cdots du_{\mu+1}\,du\, \delta(u_1)   \langle u_1|e^{ \frac{ x_1-x_2}{2\alpha^2} H}| u_{2}\rangle \cdots  \langle u_\mu| e^{ \frac{ x_\mu-x_{\mu+1}}{2\alpha^2} H}|u_{\mu+1}-v_{\nu+1}\rangle  \\ \times \delta(v_{1}-u)\langle v_1| e^{\frac{y_2-y_{1}}{2\alpha^2}H}| v_2\rangle\cdots\langle v_\nu| e^{\frac{y_{\nu+1}-y_{\nu}}{2\alpha^2}H}| v_{\nu+1}\rangle\\ \times  \Phi\big(\alpha u_1 -s_1,\ldots \alpha u_{\mu} -s_\mu; \alpha v_1-c_1,\ldots, \alpha v_{\nu}-c_\nu\big)\!\Ai(u_{\mu+1}+z)\!\Ai(u+z').
\end{multline*}
Now we make the change of variables $u_i\mapsto u_i -v_{\nu+1}$ for $i=1,\ldots, \mu$ (but
not $\mu+1$).  Because of the shift invariance of the problem we can assume without loss
of generality that $x_1=0$ (this accounts for the shift by $x_1$ in
\eqref{eq:trpr}). Since $x_{\mu+1}=0$ by assumption as well, the term that comes from the
shift of the operators telescopes to zero.  Also we introduce the function $\Ai_z(s) =
\Ai(s+z)$. We can gather the terms involving $u_{\mu+1}$ and, also including the ${\bf
  1}_{z>0}$ term, we have $\langle
u_{\mu}|e^{\frac{x_{\mu}}{2\alpha^2}H}\K|\Ai_z\rangle$. We can also gather the terms
involving $u$ along with $e^{\frac{-y_1}{2\alpha^2}z'}$ and ${\bf 1}_{z'>0}$. Observe that
$\int_{-\infty}^{\infty} du\,{\bf 1}_{z'>0}\Ai_{z'}(u) e^{\frac{-y_1}{2\alpha^2}z'}
\delta(v_1-u) = \langle \Ai_{z'}|e^{\frac{y_1}{2\alpha^2}H}\K|v_1\rangle.$ The final
result for our operator $\tilde{L}$ is
\begin{multline} \tilde{L}(z,z') = \int d\nu_1 \cdots dv_{\nu+1} du_1\cdots du_{\mu}  \delta(u_1 - v_{\nu+1}) \langle u_1|e^{ \frac{ x_1-x_2}{2\alpha^2} H}|u_2\rangle \cdots \langle u_{\mu-1}|e^{ \frac{x_{\mu-1}-x_{\mu}}{2\alpha^2} H}|u_{\mu}\rangle\\ \times \langle v_1| e^{\frac{y_2-y_{1}}{2\alpha^2}H}|v_2\rangle\cdots \langle v_\nu | e^{\frac{y_{\nu+1}-y_{\nu}}{2\alpha^2}H}|v_{\nu+1}\rangle\langle u_{\mu}|e^{\frac{x_{\mu}}{2\alpha^2}H}\K|\Ai_z\rangle \langle\Ai_{z'}|e^{\frac{y_1}{2\alpha^2}H}\K|v_1\rangle \\ \times\Phi\big(\alpha (u_1-v_{\nu+1}) -s_1,\ldots \alpha (u_{\mu}-v_{\nu+1}) -s_\mu, \alpha v_1-c_1,\ldots, \alpha v_{\nu}-c_\nu\big).\label{eq:tildeL}
\end{multline}

Recall now that we are interested in the asymptotics of this formula under the KPZ scaling
\eqref{zh}. In particular, we need to scale $(t,x)$ as $(\e^{-3/2}t,\e^{-1}x)$, which
leads to setting $\alpha=\e^{-1/2}(t/2)^{1/3}$. Observe that, with this choice,
$2\alpha^2=2^{1/3}\e^{-1}t^{2/3}$, and thus in order to obtain the desired asymptotics we
replace $x_{a} \mapsto 2^{1/3}\ep^{-1}t^{2/3}x_{a}$, $y_{a} \mapsto 2^{1/3}\e^{-1}t^{2/3}
y_{a}$, $s_{a}\mapsto\e^{-1/2}(t/2)^{1/3} s_{a}$, and $c_{a}\mapsto\e^{-1/2}(t/2)^{1/3}
c_{a}$. Note that under KPZ scaling $-c_i+y_i^2/2t$ rescales to
$(t/2)^{1/3}(-c_i+y_i^2)$ while, similarly, $s_j-x_j^2/2t$ rescales to
$(t/2)^{1/3}(-c_i+y_i^2)$. With this scaling, the left-hand side of \eqref{eq:EEkpz}
leads to
\begin{equation*}
\pp\!\left(\mathfrak{h}(t,x_k)\leq(t/2)^{1/3}(s_k-x_k^2),\,k=1\dotsc,K\right)
\end{equation*}
with $\mathfrak{h}(t,x)=T_tf(x)$ and $f(y_m)=-(t/2)^{1/3}(c_m-y_m^2)$ for $y=y_m$ and
$f=-\infty$ otherwise which, in view of \eqref{eq:trpr} is exactly what we are looking for
(recall that we have set $x_1=0$). The formula for the kernel $\bar{L}$ follows from
taking $\e\to0$, or alternatively $\alpha\to\infty$ in \eqref{eq:tildeL}. Note that
\begin{equation} \Phi\big(\alpha (u_1-v_{\nu+1} -s_1),\ldots,\alpha
  (u_{\mu}-v_{\nu+1} -s_\mu), \alpha (v_1-c_1),\ldots, \alpha (v_{\nu}-c_\nu)\big)\to {\bf
    1}_{A({\bf s,c})}\end{equation} where $A({\bf{s},\bf{c}})$ is given in (\ref{ayeaye}).
For $x_1=0$, this gives (\ref{discrtp}) after a similarity transformation. The formula for
general $x_1$ follows by simply shifting the $x$ and $y$ coordinates by $x_1$.

Finally, in order to pass to the continuum limit we first write (\ref{discrtp}) as
$\K-\int_{A^{\rm c}({\bf{s},\bf{c}})}$ where the complementary set can
be written
\begin{multline*} A^{\rm c}({\bf{s},\bf{c}}) =  \lim_{\gamma\to 0}
\bigcup_{m\in \gamma\Z}\big\{ ({\bf{u},\bf{v}}):  \max\{u_1-s_1,\ldots,  u_\mu-s_\mu\}\le m+ v_{\nu+1},\\
\min\{  c_1-v_1 ,\ldots,   c_\nu- v_\nu \} \in [m,m+\gamma)\big\}.
\end{multline*}
We obtain
\begin{align}\label{limit}
\nonumber \bar{L}(z,z') =  K-\lim_{\gamma\to 0}\sum_{m\in\gamma\Z}&
\left(\int_{B_{m,-m+\gamma}}- \int_{B_{m,-m}}\right)dv_1 \cdots dv_{\nu+1}  du_1\cdots du_{\mu} e^{(y_1-x_1) H}\K|v_1\rangle\\
&\times \langle v_1| e^{(y_2-y_{1})H}|v_2\rangle\cdots \langle v_\nu |e^{(x_{1}-y_{\nu})H}|v_{\nu+1}\rangle \delta(u_1 - v_{\nu+1})\\
\nonumber &\times \langle u_1|e^{ ( x_1-x_2) H}|u_2\rangle \cdots \cdots \langle u_{\mu-1}|e^{ (x_{\mu-1}-x_{\mu}) H}|u_{\mu}\rangle\langle u_{\mu}|e^{(x_{\mu}-x_1) H}\K,
\end{align}
where
$$
B_{m,m'} = \big\{ ({\bf{u},\bf{v}}):  \max\{u_1-s_1,\ldots,  u_\mu-s_\mu\}\le m+ v_{\nu+1},
\max\{  v_1-c_1 ,\ldots,   v_\nu- c_\nu \} \le m'\big\}.
$$
Observe that
$$
{\bf 1}_{B_{m,m'}} = \prod_{i=1}^{\nu} {\bf 1}_{u_i\le s_i+m+v_{\nu+1}} \prod_{i=1}^{\mu} {\bf 1}_{v_i \le c_i+m'}.
$$
This implies, using the fact that $\K$ is self adjoint and a projection to move it to the right entirely,
that the limit in (\ref{limit}) is
\begin{equation*}  \lim_{\gamma\to 0}\sum_{m\in\gamma\Z} \int \! dv_{\nu+1}\, \tilde\Theta^\nu_{2,m}(z',v_{\nu+1})\delta(u_1\!-\!v_{\nu+1})\tilde\Theta^\mu_{1,m,v_{\nu+1}}(u_1,z),\end{equation*}
where
\begin{equation*}
\tilde\Theta^\mu_{1,m,v_{\nu+1}}(u_1,z) = \bar{P}_{s_1+m+v_{\nu+1}} e^{( x_1-x_2) H}\cdots \bar{P}_{s_{\mu-1}+m+v_{\nu+1}} e^{ (x_{\mu-1}-x_{\mu}) H} \bar{P}_{s_\mu+m+v_{\nu+1}}e^{(x_{\mu}-x_1)H}\K,
\end{equation*} and $\tilde\Theta^\nu_{2,m} = \tilde\Theta^n_{3,-m+\gamma} - \tilde\Theta^n_{3,-m}$ where
\begin{equation*}\tilde\Theta^n_{3,m'}(z',v_{\nu+1})=e^{(y_1-x_1)H}\bar{P}_{c_1+m'} e^{( y_2-y_1) H}\cdots \bar{P}_{c_{\nu-1}+m'} e^{ (y_{\nu}-y_{\nu-1}) H} \bar{P}_{c_\nu+m'}e^{(x_{1}-y_{\nu})H}\K.
\end{equation*}

Fix two intervals $[L_s,R_s]$ and $[L_c,R_c]$, and functions $s\colon[L_s,R_s]\rightarrow \R$ and $c\colon[L_c,R_c]\rightarrow \R$. Now let
$\mu=\nu=n$ and let the $L_s\le x_{1}<\cdots< x_n\le R_s$
and $R_c\ge y_1>\cdots> y_n\ge L_c$  be evenly spaced within
these intervals.
The limit as the mesh goes to zero (i.e., $n$ goes to infinity)  in (\ref{limit}) is then given by
\begin{equation*}
\lim_{\gamma\to 0}\sum_{m\in\gamma\Z} \int dv_{\nu+1} \tilde\Theta^\infty_{2,m}(z',v_{\nu+1})\delta(u_1\! -\!v_{\nu+1})\tilde\Theta^\infty_{1,m,v_{\nu+1}}(u_1,z),
\end{equation*}
where $\tilde\Theta^\infty_{2,m} = \tilde\Theta^\infty_{3,-m+\gamma} -
\tilde\Theta^\infty_{3,-m}$ as before, and where $\tilde\Theta^\infty_{1,m,v_{\nu+1}} =
Y_{[L_s,R_s]}^{s(\cdot)+v_{\nu+1}+m} e^{(R_s-R_c) H}K$ and
$\tilde\Theta^\infty_{3,m'}=Y_{[L_c,R_c]}^{\hat{c}(\cdot)+m'} e^{(L_s-L_c)H}$ with
$Y_{[a,b]}^g$ defined in (\ref{why}) and with $\hat{c}(y) = c( L_c+R_c-y)$.  We may now
take $\gamma$ to zero. We include a multiplicative factor of $\gamma$ so that the sum
converges to an integral and use it to divide $\tilde\Theta^\infty_{2,m}$ so that the
resulting quantity converges to a derivative.
This yields (\ref{p2})-(\ref{p4}).

\appendix

\section{Two-point distribution function for the KPZ fixed point with flat initial
  condition}\label{appA}

The goal of this appendix is to obtain a formula for the two-point distribution function
of the KPZ fixed point with flat initial condition based on the  formulas proposed in Section \ref{p3}, and compare it with the two-point
distribution function for the Airy$_1$ process. In \eqref{eq:fpformula} we
need to take
\[f=f_L=0\cdot\uno{[-L,L]}+\infty\cdot\uno{[-L,L]^{\rm c}}\qquad\text{and}\qquad s(x)=
\begin{cases}
s_1 & \textrm{if } x=0,\\
s_2 & \textrm{if } x=u,\\
\infty & \textrm{otherwise}.
\end{cases}\]
Here, $L$ is a positive real number introduced because, we recall, in \eqref{eq:fpformula} $f$ is supposed to be infinite outside some compact set. Our interest
is the case $L\to\infty$.

Using these choices, what we want to compute is
\[\lim_{L\to\infty}\pp\!\left(T_2f_L(0)\leq s_0,\,T_2f_L(r)\leq
  s_1\right).\] We have chosen here to take $T_t$ with $t=2$ to simplify our computations.
Observe that in order to use the KPZ fixed point formula to compute this probability we
need to use the discrete version for the part involving $s(x)$ and the continuum version
for the flat initial condition, but one can check that this does not introduce any
difficulty.

Going back to the formula, note that in our case $L_s=0$, $R_s=r$ and
$-L_c=R_c=L$. Observe also that, since the function $c$ in \eqref{eq:fpformula} is defined
from $f$ via $f(y)=-(t/2)^{1/3}(c(y)-(y-L_s)^2)$ and we are taking $t=2$, we have $c(y)=y^2$. Therefore the KPZ fixed point
formula reads, in our case,
\begin{equation}
\pp\!\left(T_2f_L(0)\leq s_0,\,T_2f_L(r)\leq s_1\right)=\det(I-\K+G_L\K),\label{eq:kpz2}
\end{equation}
where
\[G_L(x,y)=\iint du\,dm\left[\bar P_{s_1+m+u}e^{-rH}\bar
  P_{s_2+m+u}e^{rH}\right]\!(u,x)\frac{\p}{\p
  m}\left[e^{LH}\Theta^{s^2-m}_{[-L,L]}e^{LH}\right]\!(y,u),\]
where $\Theta^{s^2-m}_{[-L,L]}$ is exactly the same operator that gives GOE in
\cite{cqr} (the exact formula will not be relevant). Note that
\[\bar P_{a_1+b}e^{-rH}\bar P_{a_2+b}e^{rH}(z_1,z_2)=\bar P_{a_1}e^{-rH}\bar P_{a_2}e^{rH}(z_1-b,z_2-b),\]
so integrating by parts we obtain
\begin{multline*}
  G_L(x,y)=\iint dm\,du\left[\delta_{s_1+m}e^{-rH}\bar
  P_{s_2+m}e^{rH}+\bar P_{s_1+m}e^{-rH}\delta_{s_2+m}e^{rH}\right]\!(0,x-u)\\
\cdot\left[e^{LH}\Theta^{s^2-m}_{[-L,L]}e^{LH}\right]\!(y,u)
\end{multline*}

The arguments in \cite{cqr} show that $\K
e^{LH}\Theta^{s^2-m}_{[-L,L]}e^{LH}\K$ converges to $\K(I-\varrho_{-m})\K$ in trace class
norm as $L\to\infty$, where
\[\varrho_a(x,y)=\delta_{x+y=2a}.\]
Hence one expects
\[\lim_{L\to\infty}\pp\!\left(T_2f_L(0)\leq s_0,\,T_2f_L(r)\leq s_1\right)=\det(I-\K-G\K),\]
with
\begin{equation*}
  G(x,y)=\iint du\,dm\left[\delta_{s_1+m}e^{-rH}\bar
    P_{s_2+m}e^{rH}+\bar P_{s_1+m}e^{-rH}\delta_{s_2+m}e^{rH}\right]\!(0,x-u)
  [I-\varrho_{-m}](y,u).
\end{equation*}

Write $G=\bar{G}-\Gamma$, where the two terms come from $I$ and $\varrho_{-m}$
in the last expression. We will first look at the term involving $I$. It is given by
\begin{align*}
  \bar{G}(x,y)&=\int du\,e^{-rH}\bar P_{s_2-s_1}e^{rH}(0,x-u)\delta_{u=y}\\
  &\hspace{0.8in}+\iint du\,dm\,\bar
  P_{s_1+m}e^{-rH}(0,s_2+m)e^{rH}(s_2+m,x-u)\delta_{u=y}\\
  &=e^{-rH}\bar P_{s_2-s_1}e^{rH}(0,x-y)+\int_{-s_1}^\infty dm
  \,e^{-rH}(0,s_2+m)e^{rH}(s_2+m,x-y).
\end{align*}
The last integral equals $e^{-rH}P_{s_2-s_1}e^{rH}(0,x-y)$, and thus $\bar G=I$. Then 
\begin{equation}
\det\!\big(I-\K+G\K)=\det\!\big(I-\Gamma\K\big).\label{eq:kpz3}
\end{equation}

Next we look at $\Gamma$. We have
\begin{multline*}
  \Gamma(x,y)=\int du\,e^{-rH}\bar P_{s_2-s_1}e^{rH}(0,x-u)\varrho_{s_1}(y,u)\\
  +\iint dm\,du\,\bar P_{s_1+m}e^{-rH}(0,s_2+m)e^{rH}(s_2+m,x-u)\varrho_{-m}(y,u).
\end{multline*}
Write $\Gamma$ as $G_1+G_2$. Then $G_1(x,y)=e^{-rH}\bar P_{s_2-s_1}e^{rH}(0,x+y-2s_1)$. By the
Baker-Campbell-Hausdorff formula (BCH) we have
\begin{align*}
  e^{-r\Delta}e^{-rH}&=e^{r^3/3}e^{r^2\nabla}e^{-r\xi}\\
  e^{rH}e^{r\Delta}&=e^{-r^3/3}e^{r\xi}e^{-r^2\nabla}.
\end{align*}
Here $\xi$ denotes the independent variable, so that $(e^{r\xi}f)(x)=e^{rx}f(x)$. Using
this we have
\begin{align*}
  G_1(x,y)&=e^{r\Delta}e^{-r\Delta}e^{-rH}\bar
  P_{s_2-s_1}e^{rH}e^{r\Delta}e^{-r\Delta}(0,x+y-2s_1)\\
  &=e^{r\Delta}e^{r^3/3}e^{r^2\nabla}e^{-r\xi}\bar
  P_{s_2-s_1}e^{-r^3/3}e^{r\xi}e^{-r^2\nabla}e^{-r\Delta}(0,x+y-2s_1)\\
  &=e^{r\Delta}\bar P_{s_2-s_1-r^2}e^{-r\Delta}(0,x+y-2s_1),
\end{align*}
where in the last equality we have used the identities $e^{-r\xi}\bar P_{a}e^{r\xi}=\bar
P_a$ and $e^{r^2\nabla}\bar P_ae^{-r^2\nabla}=\bar P_{a-r^2}$. Here, and below, we are
writing expressions involving $e^{-r\Delta}$ with $r>0$. This is justified as in
\cite{QuastelRemenikAiry1} because this operator is always applied after $B_0$ (or
$\K=B_0P_0B_0$), which is given by
\[B_0(x,y)=\Ai(x+y).\]

The kernel $G_2$ is a bit more complicated. By BCH we have
\begin{equation*}
  e^{a\nabla}e^{tH}=e^{at}e^{tH}e^{a\nabla}.
\end{equation*}
Using this we may write
\[e^{-rH}(0,s_2+m)=e^{-rH}e^{-s_2\nabla}(0,m)=e^{-s_2r}e^{-s_2\nabla}e^{-rH}(0,m)\]
and, similarly,
\begin{align*}
  e^{rH}(s_2+m,x+y+2m)&=e^{s_2\nabla}e^{rH}e^{-2m\nabla}(m,x+y)=e^{2mr}e^{(-2m+s_2)\nabla}e^{rH}(m,x+y)\\
  &=e^{2mr}e^{s_2\nabla}e^{rH}(-m,x+y)=e^{(2m+s_2)r}e^{rH}(-m,x+y-s_2),
\end{align*}
so that
\[G_2(x,y)=\int_{-s_1}^\infty dm\,e^{-rH}(-s_2,m)e^{rm}e^{rH}(-m,x+y-s_2)e^{rm}.\]
Observe that the first factor equals $e^{-rH}e^{r\xi}(-s_2,m)$ while the second one equals
$e^{-r\xi}e^{rH}(-m,x+y-s_2)$. By BCH again one has
$e^{-r\xi}e^{rH}=e^{-r\Delta}e^{-r^2\nabla}e^{-r^3/3}$ and
$e^{-rH}e^{r\xi}=e^{r\Delta}e^{r^2\nabla}e^{r^3/3}$, so using this and the symmetry of the
heat kernel above gives
\begin{align*}
  G_2(x,y)&=\int_{-s_1}^\infty dm\,e^{r\Delta}(-s_2+r^2,m)e^{-r\Delta}(-m,x+y-s_2+r^2)\\
  &=\int_{-\infty}^{s_1} dm\,e^{r\Delta}(s_2-r^2,m)e^{-r\Delta}(m,x+y-s_2+r^2)\\
  &=e^{r\Delta}\bar P_{s_1-s_2+r^2}e^{-r\Delta}(0,x+y-2s_2+2r^2).
\end{align*}

Putting the formulas for $G_1$ and $G_2$ together with \eqref{eq:kpz2} and
\eqref{eq:kpz3}, after taking $L\to\infty$, the conclusion is that
\[\pp\!\left(T_2f(0)\leq s_0,\,T_2f(r)\leq s_1\right)
=\det\!\big(I-\Gamma\K\big),\]
where
\[\Gamma(x,y)=e^{r\Delta}\bar P_{s_2-r^2-s_1}e^{-r\Delta}(0,x+y-2s_1)
+e^{r\Delta}\bar P_{s_1-s_2+r^2}e^{-r\Delta}(0,x+y-2s_2+2r^2).\]
Observe that the $r^2$ corresponds just to a parabolic shift, so writing $\tilde s_1=s_1=s_1+0^2$ and $\tilde s_2=s_2+r^2$ we get
\[\Gamma(x,y)=\Gamma_1(x,y)+\Gamma_2(x,y)=e^{r\Delta}\bar P_{\tilde s_2-\tilde s_1}e^{-r\Delta}(0,x+y-2\tilde s_1)
+e^{r\Delta}\bar P_{\tilde s_1-\tilde s_2}e^{-r\Delta}(0,x+y-2\tilde s_2).\]
This could already be considered a working formula. 

What comes next is trying to put the formula we got in a form which makes the comparison
with the Airy$_1$ formula easier. Writing $\K=B_0P_0B_0$ and using the cyclic property of
the determinant we have
\[\det(I-\Gamma K)=\det(I-P_0B_0\Gamma B_0).\]
Now
\[B_0\Gamma_1B_0(x,y)=\iint dz_1\,dz_2\,\Ai(x+z_1)e^{r\Delta}\bar P_{\tilde s_2-\tilde
  s_1}e^{-r\Delta}(0,z_1+z_2-2\tilde s_1)\Ai(z_2+y).\]
Shifting $z_1$ to $z_1-x$ and $z_2$ to $z_2-z_1+x$ 
gives
\[B_0\Gamma_1B_0(x,y)=\iint dz_1\,dz_2\,\Ai(z_1)e^{r\Delta}\bar P_{\tilde s_2-\tilde
  s_1}e^{-r\Delta}(0,z_2-2\tilde s_1)\Ai(z_2-z_1+x+y).\]
But $\int\! dz\Ai(z)\Ai(a-z)=2^{-1/3}\Ai(2^{-1/3}a)$, so letting
\[\wt B_0(x,y)=2^{-1/3}\Ai(2^{-1/3}(x+y))\]
we have deduced that
\[B_0\Gamma_1B_0(x,y)=\Gamma_1\wt B_0(0,x+y),\]
and of course the same holds with $\Gamma_2$ instead.
Letting \[\Lambda(x,y)=\Gamma\wt B_0(0,x+y)\] we have
\[\pp\!\left(T_2f(0)\leq \tilde s_0,\,T_2f(r)\leq \tilde s_1\right)
=\det\!\big(I-P_0\Lambda\big).\]

Finally we change variables $x\mapsto2^{1/3}x$, $y\mapsto2^{1/3}y$ in the
determinant. This changes the kernel $P_0\Lambda(x,y)$ to
$2^{1/3}P_0\Lambda(2^{1/3}x,2^{1/3}y)$. Writing this explicitly for the term involving
$\Gamma_1$ (and dropping the $P_0$ for a moment) gives
\begin{align*}
  2^{1/3}\int dw\,\Gamma_1(0,w)\wt B_0(w,2^{1/3}x+2^{1/3}y)&=2^{1/3}\int
  dz\, e^{r\Delta}(0,z)P_{\tilde s_2-\tilde s_1}e^{-r\Delta}\wt B_0(z,2^{1/3}x+2^{1/3}y)\\
  &=2^{2/3}\int dz\,e^{r\Delta}(0,2^{1/3}z)P_{\tilde s_2-\tilde s_1}e^{-r\Delta}\wt B_0(2^{1/3}z,2^{1/3}x+2^{1/3}y).
\end{align*}
Now $e^{r\Delta}(0,2^{1/3}z)=(4\pi
r)^{-1/2}e^{-(2^{1/3}z)^2/4r}=2^{-1/3}e^{2^{-2/3}r\Delta}(0,z)$. Likewise one can check
that $e^{-r\Delta}\wt
B_0(2^{1/3}z,2^{1/3}x+2^{1/3}y)=2^{-1/3}e^{-2^{-2/3}r\Delta}B_0(z,x+y)$. Hence the last
integral can be rewritten as $e^{2^{-2/3}r\Delta}P_{2^{-1/3}(\tilde s_2-\tilde
  s_1)}e^{-2^{-2/3}r\Delta}B_0$. The same of course holds for the term with
$\Gamma_2$. Hence the final formula becomes
\begin{equation}\label{eq:fixedptfinal}
\pp\!\left(T_2f(0)\leq s_1,\,T_2f(r)\leq s_2+r^2\right)
=\det\!\big(I-P_0\wt \Lambda\big),
\end{equation}
where $\wt\Lambda(x,y)=\wt\Gamma B_0(0,x+y)$ and
\begin{multline*}
  \wt\Gamma(x,y)=e^{2^{-2/3}r\Delta}\bar
  P_{2^{-1/3}s_2-2^{-1/3}s_1}e^{-2^{-2/3}r\Delta}(0,x+y-2^{2/3}s_1)\\
  +e^{2^{-2/3}r\Delta}\bar
  P_{2^{-1/3}s_1-2^{-1/3}s_2}e^{-2^{-2/3}r\Delta}(0,x+y-2^{2/3}s_2).
\end{multline*}

In light of the version of the Airy$_1$ formula proved in \cite{QuastelRemenikAiry1}
\begin{equation}
\pp\!\left(\aipo(0)\leq2^{-1/3}s_1,\aipo(2^{-2/3}r)\leq2^{-1/3}s_2\right)
=\det\!\big(I-B_0+\bar P_{2^{-1/3}s_1}e^{2^{-2/3}r\Delta}\bar P_{2^{-1/3}s_2}e^{-2^{-2/3}r\Delta}B_0\big)\label{eq:airy1}
\end{equation}
this suggests the conjecture
\begin{equation}
\pp\!\left(T_2f(0)\leq s_1,\,T_2f(r)\leq s_2+r^2\right)
=\pp\!\left(\aipo(0)\leq2^{-1/3}s_1,\aipo(2^{-2/3}r)\leq2^{-1/3}s_2\right).\label{eq:scaling}
\end{equation}

Unfortunately, we have not been able to check the equality of the determinants in
\eqref{eq:fixedptfinal} and \eqref{eq:airy1}, and in fact it is not at all clear whether the
equality is true. The kernels in the two formulas have many simlilarities, but observe in
particular how the variables $x,y$ appear in an odd position in
$\wt\Lambda(x,y)$. 

The formula \eqref{eq:fixedptfinal} does satisfy some basic reality checks.  The kernel
$\wt\Lambda$ is symmetric in $s_1,s_2$, which implies the same symmetry for the
two-point function. Taking $s_1\to\infty$ yields $F_{\rm GOE}(4^{1/3}s_2)$, which is the
one-point marginal of $2^{1/3}\aipo(\cdot)$. Similarly, setting $r=0$ yields $F_{\rm
  GOE}(4^{1/3}(s_1\wedge s_2))$. These three facts can be checked more or less directly
from \eqref{eq:fixedptfinal}. An additional, more complicated, reality check which can be
performed is the following (we will omit the argument, which is not hard but
involves a relatively long computation). Fix some $s\in\rr$ and let
$g(r)=\pp\big(\aipo(0)\leq2^{-1/3}s,\aipo(2^{-2/3}r)\leq2^{-1/3}s\big)$ and $\tilde g(r)=
\pp\big(T_2f(0)\leq s,\,T_2f(r)\leq s+r^2\big)$. Then $g'(0)=\tilde
g'(0)$.

We have also performed some limited numerics -- yet they
do not give a definitive answer as to the validity (or lack thereof) of this equality.

\bibliographystyle{alpha}

\end{document}